\documentclass[12pt]{article}

\usepackage{fullpage, hyperref,bbm,amssymb,amsfonts,amsthm}
\usepackage{setspace}
\usepackage{latexsym}
\usepackage{color}   
\usepackage{verbatim} 
\usepackage{graphics}
\usepackage{graphicx}
\usepackage[ table ]{ xcolor }

\input{Qcircuit}

\newtheorem{theorem}{Theorem}

\newtheorem*{problem}{Problem}

\renewcommand{\>}{\rangle}

\begin{document}
	\title{Quantum Phase Estimation with Arbitrary Constant-precision Phase Shift Operators} 
	\author {		
		Hamed Ahmadi\thanks{Department of Mathematics and Department of Electrical Engineering and Computer
		Science,
		University of Central Florida, Orlando, FL~32816, USA. Email:
		\texttt{hahmadi@cs.ucf.edu}}
		\quad			
		Chen-Fu Chiang\thanks{Department of Electrical Engineering and Computer
		Science, University of Central Florida, Orlando, FL~32816, USA.
		Email: \texttt{cchiang@eecs.ucf.edu}}
		}
\date{September 23, 2011}
	\maketitle
\begin{abstract}
While Quantum phase estimation (QPE) is at the core of many quantum algorithms known
to date, its physical implementation (algorithms based on quantum Fourier transform (QFT) ) is highly constrained by the requirement of high-precision controlled phase shift operators, which remain difficult to realize. In this paper, we introduce an alternative approach to approximately implement QPE with arbitrary constant-precision controlled phase shift operators.

The new quantum algorithm bridges the gap between QPE algorithms based on QFT and Kitaev's original approach. For approximating the eigenphase precise to the nth bit, Kitaev's original approach does not require  any controlled phase shift operator. In contrast, QPE algorithms based on QFT or approximate QFT  require controlled phase shift operators with  precision of at least Pi/2n. The new approach fills the gap and requires only arbitrary constant-precision controlled phase shift operators.
From a physical implementation viewpoint, the new algorithm outperforms Kitaev's approach.
\end{abstract}


\section{Introduction}
Quantum Phase Estimation (QPE) plays a core role in many  quantum algorithms \cite{Hallgren:02, Shor:94, Shor:05, Szegedy:04,  WCNA:09}. 
Some interesting algebraic and theoretic problems can be addressed by QPE,  such as  
prime factorization \cite{Shor:94}, discrete-log finding \cite{Shor:05},  
and order finding. 

\begin{problem}{\bf [Phase Estimation]} 
Let $U$ be a unitary matrix with eigenvalue $e^{2\pi i \varphi}$ and  corresponding eigenvector $|u\>$. Assume only a single copy of $\ket{u}$ is available, the goal is to find $\widetilde{\varphi}$ such that
\begin{equation}
\Pr(|\widetilde{\varphi}-\varphi|<\frac{1}{2^{n}})> 1-c,
\end{equation}
where $c$ is a constant less than $\frac{1}{2}$.
\end{problem}

In this paper we investigate a more general approach for the QPE algorithm. This approach completes the 
transition from Kitaev's original approach that requires no controlled phase shift operators, to QPE with approximate quantum Fourier transform (AQFT). The standard QPE algorithm utilizes the complete version of the inverse QFT. The disadvantage of the standard phase estimation algorithm is the high degree of phase shift operators required. 
Since implementing exponentially small phase shift operators is costly or physically not feasible, we need an alternative way to use lower precision operators. 
This was the motivation for AQFT being introduced --- for lowering the cost of implementation while preserving high success probability. 

In AQFT the number of required phase shift operators drops significantly with
the cost of lower success probability. Such compromise demands repeating the process extra times to achieve the final result. 
The QPE algorithm has a success probability of at least $\frac{8}{\pi^2}$ \cite{KLM:07}. Phase estimation using AQFT instead, with phase shift operators up to degree $m$ 
 where $m>\log_{2}(n) +2$,  has success probability at least $\frac{4}{\pi^2}-\frac{1}{4n}$ \cite{BEST:96, Cheung:04}.  

On the other hand, Kitaev's original approach requires only  
the first phase shift operator (as a single qubit gate not controlled). 
Comparing the existing methods, there is a gap between Kitaev's original approach and QPE with  
AQFT in terms of the degree of phase shift operators needed. In this 
paper our goal is to fill this gap and introduce a more general phase 
estimation algorithm such that it is possible to realize a phase estimation algorithm with any degree of phase shift operators in hand.
In physical implementation of the phase estimation algorithm, the depth of the circuit should be small to avoid decoherence. Also, higher degree phase shift operators are costly to implement and in many cases it is not physically feasible.

In this paper, we assume only one copy of the eigenvector $\ket{u}$ is available. This implies a restriction on the use of controlled-$U$ gates that all controlled-$U$ gates should be applied on one register. Thus, the entire process is a single circuit that can not be divided into parallel processes. 
Due to results by Griffiths and Niu, who introduced semi classical quantum Fourier transform \cite{GN:96}, quantum circuits implementing  different approaches discussed in this paper would require the same number of qubits.

The structure of this paper is organized as follows. In Sec.~\ref{KnownApproaches} we give a brief overview on existing approaches,  
such as Kitaev's original algorithm and standard phase estimation algorithm based on QFT  and AQFT. In Sec.~\ref{OurApproach} we introduce our new approach and discuss the requirements to achieve the same performance output (success probability)  as  the methods above. Finally, we make our conclusion and compare with other methods.
 

\section{Quantum phase estimation algorithms}\label{KnownApproaches}

\subsection{Kitaev's original approach}\label{sec:kitaev}
Kitaev's original approach is one of the first quantum algorithms for estimating the phase of a unitary matrix \cite{KSV:02}. Let $U$ be a unitary matrix with eigenvalue $e^{2\pi i \varphi}$ and  corresponding eigenvector $\ket{u}$ such that
\begin{equation}
U\ket{u}=e^{2\pi i \varphi}\ket{u}.
\end{equation}

In this approach, a series of Hadamard tests are performed.  In each test the phase $2^{k-1}\varphi$ ($1\leq k\leq n$) will be computed up to precision $1/16$. Assume an $n$-bit approximation is desired. Starting  from $k=n$, in each step the $k$th bit position is determined consistently from the results of previous steps.

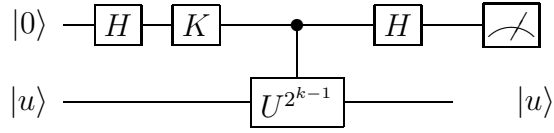
\begin{figure}
\[	\Qcircuit @C=1em @R=1em {
		\lstick{\ket{0}}   & \gate{H}  & \gate{K}     &\ctrl{1}       &\gate{H}  &\qw &\meter\\
		\lstick{\ket{u}}                                  & \qw       & \qw          &\gate{U^{2^{k-1}}} &\qw       &\qw & & \lstick{\ket{u}}   }\]
\caption{Hadamard test with extra phase shift operator.} \label{QPE_with_K_Operator}
\end{figure}

For the $k$th bit position, we perform the Hadamard test depicted in  Figure~\ref{QPE_with_K_Operator}, where  the gate $K=I_2$. Denote $\varphi_k = 2^{k-1}\varphi$, the probability of the post measurement state is
\begin{equation}\label{Eq:P1}
\Pr(0|k)= \frac{1 + \cos(2\pi \varphi_k)}{2}, \quad 
\Pr(1|k) = \frac{1 - \cos(2\pi \varphi_k)}{2}. 
\end{equation}
In order to recover $\varphi_k$, we obtain more precise estimates with
higher probabilities by iterating the process. But, this does not allow us to
distinguish between $\varphi_k$ and $-\varphi_k$. This can be solved by the same Hadamard test in Figure~\ref{QPE_with_K_Operator}, but instead we use the gate
\begin{equation}
K = \left( {\begin{array}{cc} 1 & 0  \\
 0 & i  \\
 \end{array} } \right).
\end{equation} 
The probabilities of the post-measurement states based on the modified Hadamard test become

\begin{equation}\label{Eq:P2}
\Pr(0|k)=\frac{1 - \sin(2\pi \varphi_k)}{2}, \quad 
\Pr(1|k) = \frac{1 + \sin(2\pi \varphi_k)}{2}. 
\end{equation}
Hence, we have enough information to recover $\varphi_k$ from the estimates of the probabilities.

In Kitaev's original approach, after performing the Hadamard tests, some classical post processing is also necessary.
Suppose $\varphi = 0.x_1 x_2\ldots x_n$ is an exact $n$-bit. If we are able to
determine the values of $\varphi$, $2\varphi, \ldots ,$ $2^{n-1} \varphi$ with some
constant-precision ($1/16$ to be exact), then we can determine
$\varphi$ with precision $1/2^n$ efficiently \cite{Kitaev:95, KSV:02}.

Starting with $\varphi_n$ we increase the precision of the estimated fraction as we proceed toward
$\varphi_1$. The approximated values of $\varphi_k \,(k = n, \ldots, 1)$
will allow us to make the right choices.

For $k=1,\ldots,n$ the value of $\varphi_k$ is replaced by 
$\beta_k$, where $\beta_k$ is the closest number chosen from the
set $\{\frac{0}{8},\frac{1}{8},\frac{2}{8},\frac{3}{8},\frac{4}{8},\frac{5}{8},\frac{6}{8},\frac{7}{8} \}$ 
such that
\begin{equation}\label{eqn:pa}
|\varphi_k - \beta_k|_{\text{mod 1}} <  \frac{1}{8}. 
\end{equation}

The result follows by a simple iteration. Let $\beta_n=\overline{0.x_n x_{n+1} x_{n+2}}$ and proceed by the following iteration:

\begin{equation}
x_k = \left\{ 
\begin{array}{l l}
  0 & \quad \mbox{if $\overline{0.0x_{k+1}x_{k+2}} - \beta_{k}|_{\text{mod 1}} < 1/4$ }\\ 
  1 & \quad \mbox{if $\overline{0.1x_{k+1}x_{k+2}} - \beta_{k}|_{\text{mod 1}} < 1/4$}\\ \end{array} \right. 
\end{equation}
\noindent 
for $k= n-1, \ldots, 1$. By using simple induction, the result satisfies the following inequality:
\begin{equation}
|\overline{0.x_1x_2\ldots x_{n+2}} - \varphi|_{\text{mod 1}} < 2^{-(n+2)}. 
\end{equation}

In Eq.~\ref{eqn:pa}, we do not have the exact value of $\varphi_k$. So, we have to estimate this value and use the estimate to find $\beta_k$. Let $\widetilde{\varphi_k}$ be the estimated value and
\begin{equation}
\epsilon=|\widetilde{\varphi_k}-\varphi_k|_{\text{mod 1}}
\end{equation} 
be the estimation error. Now we use the estimate to find the closest $\beta_k$. Since we know the exact binary representation of the estimate $\widetilde{\varphi_k}$, we can choose $\beta_k$ such that
\begin{equation}
|\widetilde{\varphi_k}-\beta_k|_{\text{mod 1}}\leq\frac{1}{16}.
\end{equation} 

By the triangle inequality we have,
\begin{equation}
|\varphi_k-\beta_k|_{\text{mod 1}}\leq |\widetilde{\varphi_k}-\varphi_k| _{\text{mod 1}}+ |\widetilde{\varphi_k}-\beta_k|_{\text{mod 1}}\leq\epsilon +\frac{1}{16}.
\end{equation} 

To  satisfy Eq.~\ref{eqn:pa}, we need to have $\epsilon<1/16$, which implies
\begin{equation}\label{eqn:16}
|\widetilde{\varphi_k}-\varphi_k|_{\text{mod 1}}<\frac{1}{16}.
\end{equation} 
Therefore, it is required for the phase to be estimated with precision $1/16$  at each stage.

In the first  Hadamard test (Eq.~\ref{Eq:P1}), in order to estimate $\Pr(1|k)$ an iteration of Hadamard tests should be applied  to obtain the required precision of $1/16$ for $\varphi_k$.  This is done by  counting the number of states $\ket{1}$ in the post measurement state and dividing that number by the total number of iterations performed.

The Hadamard test outputs $\ket{0}$ or $\ket{1}$ with a fixed probability. We can model an iteration of Hadamard tests as Bernoulli trials with success probability (obtaining $\ket{1}$) being $p_k$. The best estimate for the probability of obtaining the post measurement state $\ket{1}$   with $t$ samples is
\begin{equation}\label{eqn:est1}
\widetilde{p_k}=\frac{h}{t},
\end{equation} 
where $h$ is the number of ones in $t$ trials. This can be proved by Maximum Likelihood Estimation (MLE) methods \cite{HS:98}.

In order to find $\sin(2\pi\varphi_k)$ and $\cos(2\pi\varphi_k)$, we can use estimates of probabilities in Eq.~\ref{Eq:P1} and EQ.~\ref{Eq:P2}.
Let $s_k$ be the estimate of $\sin(2\pi\varphi_k)$ and $t_k$ the estimate of $\cos(2\pi\varphi_k)$. It is clear that if 
\begin{equation}
|\widetilde{p_k}-p_k|<\epsilon_0,
\end{equation} 
then   
\begin{equation}
|s_k-\sin(2\pi\varphi_k)|<2\epsilon_0,\quad
|t_k-\cos(2\pi\varphi_k)|<2\epsilon_0.
\end{equation} 

Since the inverse tangent function is more robust to error than the inverse sine or cosine functions, we use 
\begin{equation}
\widetilde{\varphi_k}=\frac{1}{2\pi}\arctan\left(\frac{s_k}{t_k}\right)
\end{equation}
as the estimation of $\varphi_k$.
By Eq.~\ref{eqn:16} we should have
\begin{equation}
\left|\varphi_k-\frac{1}{2\pi}\arctan\left(\frac{s_k}{t_k}\right)\right|_{\text{mod 1}}<\frac{1}{16}.
\end{equation}

The inverse tangent function can not distinguish between the two values $\varphi_k$ and $\varphi_k \pm 1/2$. However, because we find estimates of the sine and cosine functions as well, it is easy to determine the correct value.
The inverse tangent function is most susceptible to error when $\varphi_k$ is in the neighborhood of zero and the reason is that the derivative is maximized at zero. Thus, if 
\begin{equation}
|s_k-\sin(2\pi\varphi_k)|=\epsilon_1\quad\text{and}\quad
|t_k-\cos(2\pi\varphi_k)|=\epsilon_2,
\end{equation}
considering the case where $\varphi_k=0$, then we have
\begin{equation}
\frac{1}{2\pi}\left|\arctan\left(\frac{\epsilon_1}{1\pm\epsilon_2}\right)\right|<\frac{1}{16}.
\end{equation}
By simplifying the above inequality, we have
\begin{equation}
\left|\frac{\epsilon_1}{1\pm\epsilon_2}\right|<\tan(\frac{\pi}{8}).
\end{equation}
With the following upper bounds for $\epsilon_1$ and $\epsilon_2$, the inequality above is always satisfied when
\begin{equation}
|\epsilon_1|<1-\frac{1}{\sqrt{2}}\quad\text{and}\quad |\epsilon_2|<1-\frac{1}{\sqrt{2}}.
\end{equation}

Therefore, in order to estimate the phase $\varphi_k$ with precision $1/16$, the probabilities in Eq.~\ref{Eq:P1} and Eq.~\ref{Eq:P2} should be estimated with error at most $(2-\sqrt{2})/4$ which is approximately  0.1464.
In other words, it is necessary to find the  estimate of $\Pr(1|k)$ such that
\begin{equation}
\left|\Pr(1|k)-\frac{h}{t}\right|<\frac{2-\sqrt{2}}{4}\approx 0.1464.
\end{equation}
 
There are different ways we can guarantee an  error bound with  constant probability. The first method, used in \cite{KSV:02}, is based on the Chernoff bound. Let $X_1,\ldots,X_m$ be Bernoulli random variables, by Chernoff's bound we have
\begin{equation}\label{eqn:chernoff}
\mathrm{Pr}\left(\left|\frac{1}{m}\sum_{i=0}^{m}X_i-p_k\right|\geq \delta\right)\leq 2e^{-2\delta^2 m},
\end{equation}
where in our case the estimate is $\widetilde{p_k}=\frac{1}{m}\sum_{i=0}^{m}X_i$. Since we need an accuracy up to $0.1464$, we get
\begin{equation}\label{eqn:est2}
\mathrm{Pr}\left(|\widetilde{p_k}-p_k|> 0.1464\right)< 2e^{-(0.0429)m}.
\end{equation} 
In order to obtain 
\begin{equation}\label{eqn:est3}
\mathrm{Pr}\left(\left|\widetilde{p_k}-p_k\right|< 0.1464\right)>                   1-\frac{\varepsilon}{2},
\end{equation} 
a minimum of $m_1$ trials is sufficient when
\begin{eqnarray}\label{eqn:est4}
m_1&\approx&24\ln \frac{4}{\varepsilon}\nonumber\\
&\approx &33+24\ln \frac{1}{\varepsilon}
\end{eqnarray} 

This is the number of trials for each Hadamard test, as we have two Hadamard tests at each stage. Therefore, in order to have
\begin{equation}
\mathrm{Pr}\left(|\widetilde{\varphi_k}-\varphi_k|< \frac{1}{16}\right)> 1-\varepsilon.
\end{equation} 
we require a minimum of 
\begin{eqnarray}
m &= & 2m_1\nonumber\\
 &\approx & 47\ln \frac{4}{\varepsilon}\nonumber\\
 &\approx & 66+47\ln \frac{1}{\varepsilon}
\end{eqnarray} 
many trials.

In the analysis above, we used the Chernoff bound, which is not a tight bound. If we want to obtain the result with a high probability, we need to apply a large number of Hadamard tests. In this case, we can use an alternative method  to analyze the process by  employing methods of statistics \cite{Sivia:96}.

 Iterations of Hadamard tests have a Binomial distribution which can be approximated by a normal distribution. This  is a good approximation when $p$ is close to $1/2$ or $mp>10$ and $m(1-p)>10$, where $m$ is the number of iterations and $p$ the success probability. In other words, if we see $10$ successes and $10$ fails in our process, we can use this approximation to obtain a better bound.

In Kitaev's algorithm each Hadamard test has to be repeated a sufficient number of times to achieve the 
required accuracy with high probability. 
Because only one copy of $\ket{u}$ is available, all controlled-$U$ gates have to be applied to one register. Therefore, 
all the Hadamard tests have to be performed in sequence, instead of parallel,  during one run of the
circuit. A good example for this case  is the order finding algorithm. We refer
the reader to \cite{NC:00} for more details.

In Kitaev's approach, there are $n$ different Hadamard 
tests that should be performed. 
Thus, if the probability of error in each Hadamard test is $\varepsilon_0$, by applying the union bound, the error probability of the entire process is $\varepsilon=n\varepsilon_0$. Therefore, in order to obtain
\begin{equation}
\Pr(|\varphi-\widetilde{\varphi}|<\frac{1}{2^n})> 1-\varepsilon,
\end{equation}
for approximating each bit we need $m$ trials where
\begin{equation}\label{m:kitaev}
m =47\ln \frac{4n}{\varepsilon}.
\end{equation} 
Since, all of these trials have to be done in one
circuit, the circuit consists of
$mn$ Hadamard tests. Therefore the circuit involves $mn$  controlled-$U^{2^k}$ operations. As a result, if a constant success probability is desired, the depth of the circuit will be $O(n\log n)$.

\subsection{Approach based on QFT}\label{StandardPE}
\begin{figure}
\[\scalebox{1}{
	\Qcircuit @C=.7em @R=0.7em {
	\lstick{|0\>}      & \qw    & \gate{H}    & \qw            & \qw            & \qw    &  &     & \qw & \ctrl{4}           & \qw & \multigate{3}{{\rm QFT}^\dagger} & \qw \\
	                   &        &             & \vdots         &                &        & \cdots &     &     &                &     &        &  \\
	\lstick{|0\>}      & \qw    & \gate{H}    & \qw            & \ctrl{2}       & \qw    &  &     & \qw & \qw                & \qw & \ghost{{\rm QFT}^\dagger}        & \qw \\ 
	\lstick{|0\>}      & \qw    & \gate{H}    & \ctrl{1}       & \qw            & \qw    &  &     & \qw & \qw                & \qw & \ghost{{\rm QFT}^\dagger}        & \qw \\
	\lstick{|u\>}      & \qw    & \qw         & \gate{U^{2^0}} & \gate{U^{2^1}} & \qw    &  &     & \qw & \gate{U^{2^{n-1}}} & \qw & \qw                      & \qw } }\]
 \caption {Standard Quantum Phase Estimation.} 
 \label{QPEfig}
\end{figure}
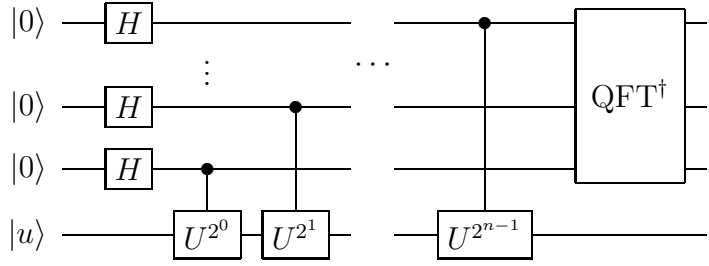

One of the standard methods to approximate the phase of a unitary matrix is QPE based on QFT. The structure of this method
is depicted at Figure ~\ref{QPEfig}. The QPE algorithm 
requires two registers and contains two stages. If an $n$-bit approximation of the phase $\varphi$ is desired, then  the 
first register is  prepared as a composition of $n$ qubits initialized in the state $|0\>$. The second register is initially prepared in the state $\ket{u}$. 
The first stage prepares a uniform superposition over all possible states and
then applies controlled-$U^{2^k}$  operations. Consequently, the state will become
\begin{equation}\label{stateStage1}
\frac{1}{2^{n/2}}\sum_{k=0}^{2^n-1}e^{2 \pi i \varphi k}|k\>. 
\end{equation}
The second stage in the QPE algorithm is the QFT$^\dag$ operation. 
  
There are different ways to interpret the inverse Fourier transform. In the QPE
algorithm, the post-measurement state of each qubit in the first register represents a bit in the final approximated binary fraction of the phase. Therefore, we can consider computing each bit as a step.
The inverse Fourier transform can be interpreted such that at each step (starting from the least significant bit),  using the 
information from previous steps, it transforms the state 
\begin{equation}
\frac{1}{\sqrt{2}}(\ket{0}+e^{2\pi i 2^k\varphi}\ket{1})
\end{equation}
to get closer to one of the states 
\begin{eqnarray}\frac{1}{\sqrt{2}}(\ket{0}+e^{2\pi i 0.0}\ket{1})&=&\frac{1}{\sqrt{2}}(\ket{0}+\ket{1}) \nonumber\\
&\text{or}&\notag\\
\frac{1}{\sqrt{2}}(\ket{0}+e^{2\pi i 0.1}\ket{1})&=& \frac{1}{\sqrt{2}}(\ket{0}-\ket{1}). 
\end{eqnarray}

Assume we are at step $k$ in the first stage. By applying  controlled-$U^{2^k}$
operators due to phase kick back,  we obtain the state
\begin{equation}\label{state1}
\frac{\ket{0}+e^{2\pi i 0.x_{k+1}x_{k+2}\ldots x_n}\ket{1}}{\sqrt{2}}.
\end{equation}

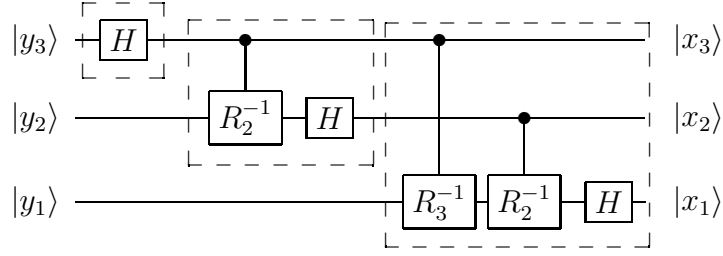
\begin{figure}
\[\scalebox{1}{
	\Qcircuit @C=.4em @R=1em {  
	       \lstick{|y_3\>} &\qw      &\gate{H} &\qw &\qw  &\qw &\qw &\ctrl{1}  &\qw &\qw     &\qw &\qw      &\qw &\ctrl{2}        &\qw             &\qw      &\qw&\qw & \rstick{|x_3\>} \\ 
	       \lstick{|y_2\>} &\qw   &\qw   &\qw   &\qw      &\qw&\qw &\gate{R_2^{-1}}    &\qw &\gate{H} &\qw &\qw      &\qw &\qw  &\ctrl{1}        &\qw &\qw      &\qw & \rstick{|x_2\>} \\ 
	       \lstick{|y_1\>} &\qw    &\qw  &\qw    &\qw     &\qw &\qw &\qw                &\qw &\qw      &\qw &\qw &\qw&\gate{R_3^{-1}} &\gate{R_2^{-1}} &\qw &\gate{H} &\qw & \rstick{|x_1\>}  
 	        \gategroup{1}{3}{1}{3}{1.1em}{--}
 	        \gategroup{1}{8}{2}{10}{1.3em}{--}
 	        \gategroup{1}{14}{3}{17}{1.1em}{--}
	        }	
}	        \]

\caption{3-qubit inverse QFT where $1 \leq i \leq 3$,
$|y_i\>=\frac{1}{\sqrt{2}}(\left|0\right>+e^{2\pi i(0.x_i\ldots x_3)}\left|1\right>$). }
  \label{fig:InverseQFT_3qubit}	        
\end{figure}

Shown in Figure ~\ref{fig:InverseQFT_3qubit}, each step (dashed-line box) uses
the result of previous steps, where  phase shift operators are defined as
\begin{equation}\label{eq:phaseshift}
R_k \equiv
\left[ {\begin{array}{cc}
 1 & 0  \\
 0 & e^{2\pi i/2^k}  \\
 \end{array} } \right] 
\end{equation}
 for $2\leq k \leq n$.

By using the previously determined bits $x_{k+2},\ldots, x_n$ and the action of corresponding controlled phase shift operators (as depicted in Figure~\ref{fig:InverseQFT_3qubit}) the state in Eq.~\ref{state1} becomes  
\begin{equation}
\frac{\ket{0}+e^{2\pi i 0.x_{k+1}0\ldots 0}\ket{1}}{\sqrt{2}}=\frac{\ket{0}+(-1)^{x_{k+1}}\ket{1}}{\sqrt{2}}.
\end{equation}
Thus, by applying a Hadamard gate to the state above we obtain $\ket{x_{k+1}}$. Therefore, we can consider the inverse Fourier 
transform as a series of Hadamard tests.

If $\varphi$ has an exact $n$-bit binary representation the success probability at each step is $1$. While, in the  case that $\varphi$ cannot be
exactly expressed in $n$-bit binary fraction, the success probability $P$ of the post-measurement state, at step $k$, is 
\begin{equation}
P=\cos^2(\pi \theta) \quad \text{for} \quad |\theta|<\frac{1}{2^{k+1}}
\end{equation} 
Detailed analysis obtaining similar probabilities are given in Sec.~\ref{OurApproach}. 

Therefore, the success probability increases as we proceed. The following theorem gives us the success probability of the QFT algorithm.

\begin{theorem}[\cite{KLM:07}]\label{PElowerbound}
If $\frac{x}{2^n} \le \varphi \le \frac{x+1}{2^n}$, then
the phase estimation algorithm returns one of $x$ or $x+1$ with probability at least
$\frac{8}{\pi^2}$.
\end{theorem}


\subsection{Approach based on AQFT}\label{sec:AQFT}

\begin{figure}[t]
\[\scalebox{1}{
  \Qcircuit @C=0.7em @R=0.7em {
\lstick{|x_1\>}      & \gate{H}  & \gate{R_2}     & \qw    &\cdots  &        &\gate{R_{m-1}} & \qw    & \gate{R_m} &\qw      &\qw    &\qw           &\qw             &\qw            &\qw             &\qw        &\qw        &\qw       &\qw     &&&     &\lstick{|y_1\>}                                                  \\ 
\lstick{|x_2\>}      & \qw       & \ctrl{-1}      & \qw    &\qw     & \qw    & \qw           & \qw    & \qw        &\gate{H} &\qw    &\cdots        &                &\gate{R_{m-1}} &\gate{R_{m}}    &\qw        &\qw        &\qw       &\qw     &&&     &\lstick{|y_2\>}                       \\
	                     &           & \vdots         &        &\cdots  &        &\ctrl{-2}      & \qw    &  \cdots    &         &       &              &                &               &                &           &           &          &        &&&     &\\ 
	                     &           & \vdots         &        &        &        &               &        &  \ctrl{-3} & \qw     &\qw    &\cdots        &                & \ctrl{-2}     & \qw            &\qw        & \cdots    &          &        &&&     &\\ 
	                     &           & \vdots         &        &        &        &               &        &            &         &       &              & \cdots         &               & \ctrl{-3}      &\qw        &\cdots     &          &        &&&     &\\ 
\lstick{|x_{n-1}\>}  & \qw       & \qw            & \qw    &\cdots  &        & \qw           & \qw    & \qw        &\qw      &\qw    &\qw           &\qw             &\qw            &\qw             &\gate{H}   &\gate{R_2} &\qw       &\qw     &&&&     &\lstick{|y_{n-1}\>}                          \\
\lstick{|x_n\>}      & \qw       & \qw            & \qw    &\cdots  &        & \qw           & \qw    & \qw        &\qw      &\qw    &\qw           &\qw             &\qw            &\qw             &\qw        &\ctrl{-1}  &\gate{H}  &\qw     &&&     &\lstick{|y_n\>}                        }
  }\]
  \caption{Quantum circuit for AQFT.}
  \label{AQFT}
\end{figure}
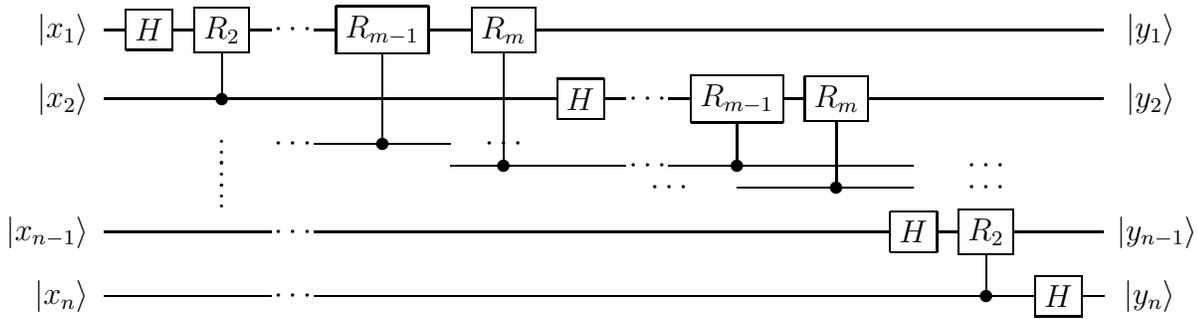

AQFT was first introduced by Barenco, et al \cite{BEST:96}. It has the 
advantage in algorithms that involve periodicity estimation. Its structure is
similar to regular QFT but differs by eliminating higher precision 
phase shift operators. The circuit of AQFT is shown in Figure~\ref{AQFT}.  At the RHS of the 
circuit, for $ n-m < i \leq n$
\begin{equation}
\ket{y_i} =
 \frac{1}{\sqrt{2}}(\left|0\right>+e^{2\pi i(0.x_i\ldots x_n)}\left|1\right>)  \end{equation}
and for $1<i \leq n-m$,
\begin{equation}
\ket{y_i}= \frac{1}{\sqrt{2}} (\left|0\right>+e^{2\pi i(0.x_i\ldots x_{i+m-1})}\left|1\right>). 
\end{equation}

Let $0.x_1x_2\ldots x_n$ be the binary representation of 
eigenphase $\varphi$. For estimating each $x_p$, where $1 \leq p \leq n$,
AQFT$_m$ requires at most $m$ phase shift operations. Here $m$ is defined as the degree 
of the AQFT$_m$. 

Therefore, phase shift operations in AQFT$_m$ requires precision up to   $e^{2 \pi i /2^m}$. 
The  probability $P$ of gaining an accurate output using AQFT$_m$, when $ m \geq \log_2 n + 2$, is at least \cite{BEST:96}  
\begin{equation}\label{eq1}
P \geq \frac{8}{\pi^2}(\sin^2(\frac{\pi}{4}\frac{m}{n})).
\end{equation}

The accuracy of AQFT$_{m}$ approaches the lower bound  for the accuracy of 
the full QFT, which is $\frac{8}{\pi^2}$. A better
lower bound is also achieved by Cheung in \cite{Cheung:04}
\begin{equation}
P \geq \frac{4}{\pi^2}-\frac{1}{4n}.
\end{equation}

Moreover, this indicates the logarithmic-depth AQFT provides an 
alternative approach to replace the regular QFT in many
quantum algorithms. The total number of the phase shift operator invocations in AQFT$_m$ is $O(n \log_2 n)$,
instead of $O(n^2)$ in the QFT. The phase shift operator precision requirement is only up to 
$e^{2 \pi i /4n}$, instead of $e^{2 \pi i /2^n}$.

By using the AQFT instead of the QFT we trade off smaller success probability with smaller degrees of phase shift 
operators and a shorter circuit.


\section{New approach with constant degree phase shift operators }\label{OurApproach}

 In this section we introduce our new approach for QPE. Our approach draws a trade-off between the highest degree of phase shift 
 operators being used and the depth of the circuit. As a result, when smaller
 degrees of phase shift operators are used,  the depth of the circuit increases and vice versa.
 
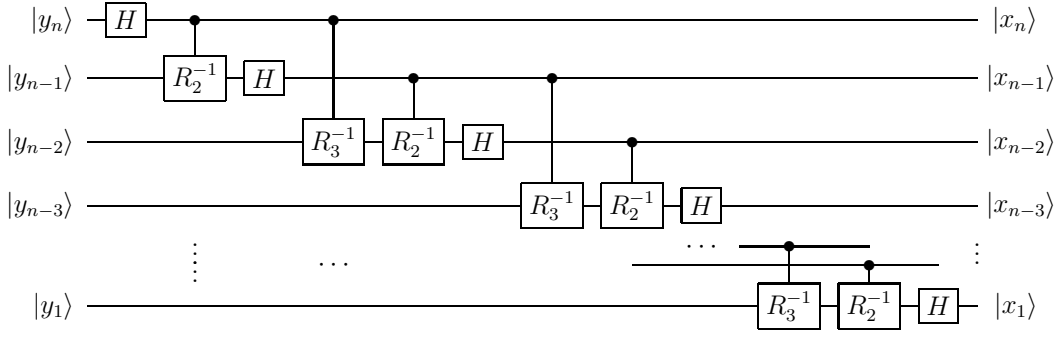
\begin{figure}
 \[\scalebox{0.85}{
   \Qcircuit @C=0.7em @R=0.7em {
 \lstick{|y_n\>}        &\gate{H}   &\ctrl{1}         &\qw             &\ctrl{2}            &\qw                 &\qw         &\qw                 & \qw                & \qw                 & \qw    &\qw                 & \qw                & \qw     &\qw &&&       &\lstick{|x_n\>}                                                  \\ 
 \lstick{|y_{n-1}\>}    &\qw        &\gate{R^{-1}_2}  &\gate{H}        &\qw                 &\ctrl{1}            &\qw         &\ctrl{2}            &\qw                 & \qw                 & \qw    & \qw                & \qw                & \qw     &\qw &&&&     &\lstick{|x_{n-1}\>}                       \\
 \lstick{|y_{n-2}\>}    &\qw        &\qw              & \qw            &\gate{R^{-1}_{3}}   &\gate{R^{-1}_{2}}   &\gate{H}    & \qw                & \ctrl{1}           & \qw                 & \qw    &\qw                 & \qw                & \qw     &\qw &&&&     &\lstick{|x_{n-2}\>}\\ 
 \lstick{|y_{n-3}\>}    &\qw        &\qw              &\qw             & \qw                &  \qw               & \qw        &\gate{R^{-1}_{3}}   &\gate{R^{-1}_{2}}   &\gate{H}             &\qw     & \qw                &\qw                 & \qw     &\qw &&&&     &\lstick{|x_{n-3}\>}\\ 
                        &           & \vdots          &                &                    &                    &            &                    &                    &\cdots               &        & \ctrl{2}           & \qw                &         &\vdots             &  \\ 
                        &           &\vdots           &                & \cdots             &                    &            &                    &                    &\qw                  &\qw     &\qw                 &\ctrl{1}            &\qw      &    &&&       &                            \\ 
  \lstick{|y_{1}\>}      &\qw        &\qw              &\qw             &\qw                 & \qw                & \qw        & \qw                & \qw                &\qw                  &\qw     &\gate{R^{-1}_{3}}   &\gate{R^{-1}_{2}}   &\gate{H} &\qw &&&        &\lstick{|x_1\>}                        }
 }\]
 
   \caption{QPE with only two controlled phase shift operations.}
   \label{QFT_Ours}
 \end{figure}

As pointed out in Sec.~\ref{StandardPE}, by using information of previous qubits, the full-fledged inverse QFT transforms the phase such that the phase of the corresponding qubit gets closer to one of the states $\ket{+}$ or $\ket{-}$. For our approach, we first  consider the case where only the controlled phase shifts operators $R_2$ and
 $R_3$ are used (Eq.~\ref{eq:phaseshift}). In this case, we only use the information of the two previous qubits (see Figure~\ref{QFT_Ours}).
 In such a setting, we show that it is possible to perform the QPE algorithm with arbitrary
 success probability.  
 
 The first stage of our algorithm is similar to the first stage of QPE based on QFT.
 Assume the phase is $\varphi=0.x_1x_2x_3\ldots$ with an infinite
 binary representation.  At step $k$, the phase after the action of the controlled gate $U^{2^k}$ is $2^k \varphi=0.x_{k+1}x_{k+2}\ldots$ and the corresponding state is
 \begin{equation}\ket{\psi_k}=\frac{1}{\sqrt{2}}(\ket{0}+e^{2\pi i 2^k\varphi}\ket{1}).\end{equation} 
 
 By applying controlled phase shift operators $R_2$ (controlled by the $(k-1)$th qubit)  and $R_3$ (controlled by the $(k-2)$th qubit) to the state above, we obtain
 \begin{equation}\ket{\widetilde{\psi_k}}=\frac{1}{\sqrt{2}}(\ket{0}+e^{2\pi i \widetilde{\varphi}}\ket{1}),\end{equation}
where 
\begin{equation}\widetilde{\varphi}=0.x_{k+1}00x_{k+4}\ldots.\end{equation}
  
It is easy to see that
 \begin{equation}|\widetilde{\varphi}-0.x_{k+1}|<\frac{1}{8}.\end{equation}
 Hence, we can express
 \begin{equation}\widetilde{\varphi}=0.x_{k+1}+\theta\end{equation}
 where $|\theta|<\frac{1}{8}$. Therefore, the state  $\ket{\widetilde{\psi_k}}$ can be rewritten as
 \begin{equation}\ket{\widetilde{\psi_k}}=\frac{1}{\sqrt{2}}(\ket{0}+e^{2\pi i (0.x_{k+1}+\theta)}\ket{1}).\end{equation}
 
In order to approximate the phase $\varphi$ at this stage ($k$th step), we need to find the value of $x_{k+1}$ by measuring the $k$th qubit. In this regard, we first apply a Hadamard gate before the measurement to the state $\ket{\widetilde{\psi_k}}$. The post-measurement state will determine the value of $x_{k+1}$ correctly with high probability. The post measurement probabilities of achieving $\ket{0}$ or $\ket{1}$ in the case where $x_{k+1}=0$ is
 \begin{eqnarray}
 \Pr(0|k)&=&\cos^2(\pi\theta)\nonumber \\
 \Pr(1|k)&=&\sin^2(\pi\theta). 
 \end{eqnarray} 
 Therefore,
 \begin{eqnarray}
 \Pr(0|k)&\geq&\cos^2(\frac{\pi}{8} )\approx 0.85 \nonumber \\
 \Pr(1|k)&\leq&\sin^2(\frac{\pi}{8} )\approx 0.15 
 \end{eqnarray} 
 In the case where $x_{k+1}=1$, the success probability is similar. 
 
 By iterating this process a sufficient number of times and then letting the majority decide, we can achieve any desired accuracy. 
The analysis is similar to  Sec.~\ref{sec:kitaev}. In this case, all we require is to find the majority. Therefore, by a simple application of the Chernoff's bound 
\begin{equation}
\mathrm{Pr}\left(\frac{1}{m}\sum_{i=0}^{m}X_i\leq \frac{1}{2}\right)\leq e^{-2m(p-\frac{1}{2})^2 },
\end{equation}
where in this case $p=\cos^2(\pi/8)$. It is easy to see that if a success probability of $1-\varepsilon$ is required, then 
we need at least
\begin{equation}\label{m:ours}
m=4\ln(\frac{1}{\varepsilon})
\end{equation}
many trials for approximating each bit.

%
%
%

\begin{figure}[ht]
\begin{center}
\includegraphics[scale=1.4]{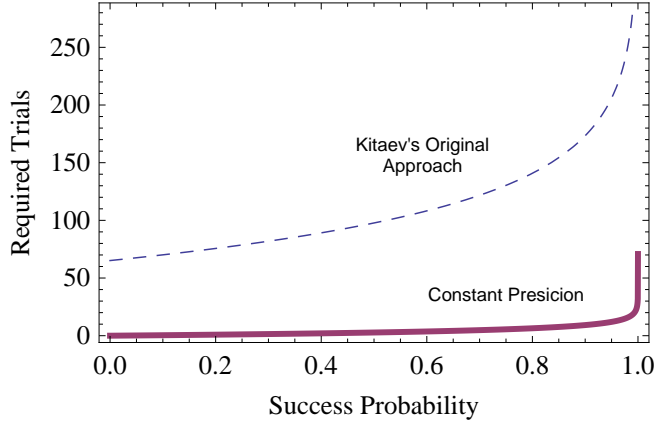}
\caption{Required trails for estimating each bit in Kitaev's original approach and our new approach.}\label{KitaevVSOurs}
\end{center}
\end{figure}

By comparing Eq.~\ref{m:kitaev} and Eq.~\ref{m:ours} (Table~\ref{table:3}), we see that while preserving the success probability,  our new algorithm  differs by a constant and scales about 12 times better than Kitaev's original approach in terms of the number of Hadamard tests required (Figure \ref{KitaevVSOurs}). 
In physical implementations this is very important, especially in the case where only one copy of the eigenvector $\ket{u}$ is available and all Hadamard tests should be performed during one run of the circuit. 
  
 In the algorithm introduced above, only phase shift operators $R_2$ and $R_3$ are used. When higher phase shift operators are used in our algorithm, the success probability of each Hadamard test will increase. As a result, 
 fewer trials are required in order to achieve similar success probabilities.  
 As pointed out in Sec.~\ref{sec:AQFT}, the QPE based on AQFT requires phase shift operators of degree at least $2 +\log n$. With this precision of phase shift operators in hand, the success probability at each step would be high enough such that there is no need to iterate each step. In such scenario, one trial is sufficient to achieve an overall success probability of a constant. 
 
\begin{table}[ht]
\begin{center}
\begin{tabular}{ |c | c|c| } 
\hline
 {\bf Success} &{\bf Kitaev's} &{\bf Constant}\\
{\bf Probability} &{\bf Original Approach} & {\bf Precision}\\ \hline 
0.50000 &  98& 3 \\ \hline
0.68269 &   120 & 5 \\ \hline
0.95450& 211 & 13 \\ \hline
0.99730& 344 & 24 \\ \hline 
0.99993 &515 & 39\\ \hline
\end{tabular}
\caption{Required trials for estimating each bit by using Chernoff's bound. }\label{table:3}
\end{center}
\end{table}

Recall the phase estimation problem stated in the introduction. If  a constant success probability greater than $\frac{1}{2}$ is required, the depth of the circuit for all the methods mentioned in this paper (except the QPE based on full fledged QFT, which is $O(n^2)$), would be $O(n\log n)$ (assuming the cost of implementing the controlled-$U^{2^k}$ gates are all the same). This means the depth of the circuits differ only by a constant. However, the disadvantage of Kitaev's original approach to our new approach is the large number of  Hadamard tests required for each  bit in the approximated fraction. 

Therefore, the new method introduced in this paper  provides  the flexibility of using any available degree of controlled phase shift operators while preserving the success probability and the length of the circuit up to a constant.

\label{conclusions}	
\section{Acknowledgments}
		We would like to thank Pawel Wocjan for useful discussions and Stephen Fulwider for helpful comments. H.~A. and C.~C. gratefully acknowledge the support of NSF grants	CCF-0726771 and CCF-0746600. \\
		

\end{document}